\def\fun#1#2{\lower3.6pt\vbox{\baselineskip0pt\lineskip.9pt
  \ialign{$\mathsurround=0pt#1\hfil##\hfil$\crcr#2\crcr\sim\crcr}}}
\relax
\documentstyle[12pt]{article}
\advance\textheight by \topskip
\begin{document}
\thispagestyle{empty}

\rightline{LANDAU--91--TMP--1}
\rightline{May, 1991}
\rightline{Revised: October, 1991}

\begin{center}
\large
{\LARGE STOCHASTIC INFLATION: NEW RESULTS\footnote{\normalsize Talk,
presented
at
the First International Sakharov Conference in Physics, Moscow, May
21 -- 31,
1991.
To appear in the proceedings (Nova Science Pub., New York).} }  \\
\vspace{1cm}
{\bf A.Mezhlumian \hspace{0.3cm} and \hspace{0.3cm}
A.A.Starobinsky}\\
\vspace{1cm}
{\it L.D.Landau Institute for
Theoretical Physics,\\
Russian Academy of Sciences, \\
Kosygina st.\ 2, Moscow, 117334, Russia} \\
\end{center}

\vspace{1cm}
{\centerline {\bf Abstract}}

We prove that, in stochastic approach, there exists an equivalence
relation between different inflationary models
under some redefinition of field and time variables.
The postinflationary physics is insensitive to it and one can say
that
related theories are in fact indistinguishable from the viewpoint
of a local observer. We discuss the methodology of cosmology
as a branch of natural science and present a general prescription
for the interactive development of cosmological
theory and observations avoiding the problems of freedom in the
choice
of different variants of the theory and unobservability of
superhorizon scales. Also we present all ``exactly solvable''
inflationary models in slow rolling regime.

\vspace{2cm}


\newcommand{\tV}[2]{\mbox{$\mbox{$\tilde{V}$}^{#1}_{#2}({\cal F})$}}
\newcommand{\VP}[2]{\mbox{$V^{#1}_{#2}(\varphi ({\cal F}))$}}
\newcommand{\Vp}[2]{\mbox{$V^{#1}_{#2}(\varphi $)}}
\newcommand{\tVs}[2]{\mbox{$\mbox{$\tilde{V'}$}^{#1}_{#2}({\cal
F})$}}
\newcommand{\VPs}[2]{\mbox{$V'^{#1}_{#2}(\varphi ({\cal F}))$}}
\newcommand{\Vps}[2]{\mbox{$V'^{#1}_{#2}(\varphi )$}}
\newcommand{\Te}[1]{\mbox{$T^{#1}(\varphi )$}}
\newcommand{\Fe}[1]{\mbox{$F^{#1}(\varphi )$}}
\newcommand{\FE}[1]{\mbox{$F^{#1}(\varphi ({\cal F}))$}}
\newcommand{\TE}[1]{\mbox{$T^{#1}(\varphi ({\cal F}))$}}
\newcommand{\kep}[1]{\mbox{$\exp{(#1 \alpha \varphi )}$}}
\newcommand{\ah}{\mbox{$\frac{1}{2} $}}
\newcommand{\pr}[1]{\mbox{$\propto \varphi^{#1} $}}
\newcommand{\pro}[1]{\mbox{$\propto #1 $}}
\newcommand{\Xl}[2]{\mbox{$\frac{\mbox{$#1$}}{#2} \, \varphi^{#2} $}}
\newcommand{\XL}[2]{\mbox{$\frac{\mbox{$#1$}}{#2} \, {\cal F}^{#2}
$}}

\newpage

Stochastic approach \cite{r1}
to inflationary models (see \cite{r2}
for review) became the most popular and powerful, especially
in the case of new \cite{r3}
and chaotic \cite{r4}
scenarios. However, there are still some mysterious coincidences
in various calculations in the framework of new and chaotic
scenarios.
For example, it is well known, that primordial perturbations spectra
in some new and chaotic models are the same (compare \cite{r5}
and \cite{r2} ) \@. It was proposed some time ago \cite{r6}
that there is an underlying equivalence between the corresponding
models. We investigate this equivalence in detail in the present
work.

In the slow rolling approximation ($\dot{H} \ll H^{2} $),
dynamics of the inflaton field $\varphi $
is described by the stochastic Langevin equation \cite{r1}
\begin{equation}
   \frac{d \varphi}{dt} = -\frac{\Vps{}{\varphi }}{3H(\varphi )} +
           \frac{H^{3/2}(\varphi )}{2 \pi} \, \xi (t)
\label{eq1}
\end{equation}
where $\xi (t)$
is the gaussian white noise
\begin{equation}
  < \xi (t_{1}) \, \xi (t_{2})> = \delta (t_{1} \, -t_{2})
\label{eq2}
\end{equation}

We shall consider (\ref{eq1})
in the sense of Stratonovich, rather then It\^{o} , because $\xi (t)$
should be treated as a limit of  colored  noise  when
the decomposition of $\varphi $ into
short- and long-wavelength modes in the coarse-graining procedure
\cite{r1}
becomes sharp. The other important point is that the stochastic
term in (\ref{eq1}) is written
as a multiplicative, rather then additive noise. This is dictated by
independence of the short-wavelength mode phase on the value of
coarse-grained field $\varphi $\@.
This corresponds to the following choice of the operator ordering in
the Fokker-Planck equation:

\begin{equation}
\frac{\partial}{\partial t} \, P(\varphi ,t) =
\frac{\partial }{\partial \varphi } \, \left(
\frac{\Vps{}{\varphi }}{3H(\varphi )} \, P(\varphi ,t) +
\frac{1}{2} \, \frac{H^{3/2}(\varphi )}{2 \pi} \,
\frac{\partial }{\partial \varphi } \, \left(
\frac{H^{3/2}(\varphi )}{2 \pi} \, P(\varphi ,t) \right) \right)
\label{eq2.1}
\end{equation}
Let us rewrite (\ref{eq1})
in the form
\begin{equation}
   \frac{d \varphi}{dt} = -A \frac{\Vps{}{\varphi }}{\Vp{1/2}{}}
   + B \Vp{3/4}{} \, \xi (t)                      \label{eq3}
\end{equation}
and make the following changes:
\begin{description}
\item[i)]          change of the inflaton field variable
   \begin{equation}
         \varphi \rightarrow {\cal F}(\varphi) = \int^{\varphi} d
\chi \,
          F(\chi)                                 \label{eq4}
   \end{equation}
\item[ii)]         change of the time variable
   \begin{equation}
          t \rightarrow \tau (t) = \int^{t} ds \, T(\varphi (s))
\label{eq5}
   \end{equation}
\end{description}
Here $T(\varphi (s))$ is the function of $\varphi (s) $
which is a solution of (\ref{eq3}) \@.
In fact, this is the most general form of time reparametrization
consistent with inflationary ansatz (during inflation everything
may depend only on a local value of coarse-grained inflaton field)\@.
The only restriction is $\Te{} > 0 $ \@.
Thus, we have
\begin{equation}
\frac{d{\cal F}}{d\tau}=-A\frac{\Fe{} \, \Vps{}{\varphi}}{\Vp{1/2}{}
\, \Te{}}
  +B \frac{\Fe{} \, \Vp{3/4}{}}{\Te{1/2}} \, \eta (\tau)  \label{eq6}
\end{equation}
\begin{equation}
  <\eta (\tau_{1})\,\eta (\tau_{2})>=\delta (\tau_{1} \, -\tau_{2})
\label{eq7}
\end{equation}
where \Te{-1/2} in the last term of (\ref{eq6})
is due to the redefinition of noise $\eta (\tau) $ (\ref{eq7}) \@.
One can try to find a new potential \tV{}{}
corresponding to $V,F$ and $T$ in order to rewrite (\ref{eq6})
in the initial form (\ref{eq3}) \@.
It would mean that in the slow rolling approximation a theory
with \tV{}{} is equivalent to the theory with $V(\varphi )$ \@.
\begin{equation}
  \frac{\FE{2} \, \VPs{}{{\cal F}}}{\VP{1/2}{} \, \TE{}} =
  \frac{\tVs{}{\cal F}}{\tV{1/2}{}}                \label{eq8a}
\end{equation}
\begin{equation}
  \frac{\FE{} \, \VP{3/4}{}}{\TE{1/2}}=\tV{3/4}{}       \label{eq8b}
\end{equation}
It is easy to derive from Eqs.\ (\ref{eq8a}) ,(\ref{eq8b}) that
\begin{equation}
   \frac{1}{\tV{}{}}=\frac{1}{\VP{}{} }+\frac{1}{V_{0}}  \label{eq9}
\end{equation}
\begin{equation}
   \Te{}=\Fe{2} \left( 1+\frac{\Vp{}{}}{V_{0}} \right)^{3/2}
\label{eq10}
\end{equation}

Let us make some comments on this result. The Eq.\  (\ref{eq9})
tells us what the new theory \tV{}{}
corresponds to old one $V(\varphi )$
given a function $F(\varphi )$ \@.
At the same time Eq.\  (\ref{eq10})
shows for what time parameterization does this equivalence
hold (note that $T>0$ in this case)\@.
One can also interpret Eqs.\  (\ref{eq9}) ,(\ref{eq10})
in another way, that is, what class of theories \tV{}{}
corresponds to the old $V(\varphi )$
given a time parameterization $T(\varphi )$ \@.
In this case Eq.\  (\ref{eq10})
determines the function $F(\varphi )$
and then Eq.\  (\ref{eq9})
gives \tV{}{} \@.

A remarkable feature of (\ref{eq9})
is that in many cases it relates chaotic models to new inflationary
ones.
Often (\ref{eq9})
can be written as
\begin{equation}
  \tV{}{} \approx V_{0} - \frac{V^{2}_{0}}{\VP{}{}}    \label{eq11}
\end{equation}
where $\VP{}{} >> V_{0}$
in the region of chaotic dynamics of inflaton.

One can easily see from (\ref{eq9}),(\ref{eq10}) that in the case of
$R+R^{2}$ gravity, when the effective scalaron potential
$\Vp{}{} \rightarrow {\it constant}$ for
$\varphi \rightarrow \infty $ \cite{St80},
$T=1$ variable changing leaves
the scalaron potential unchanged. One can say that the
model \cite{St80} is form invariant with respect to our
equivalence transformation. May be this is related with some
hidden symmetry of this model. We don't attempt to
explore this question here but, in principle, it is a very
interesting direction of future research.

{}From the viewpoint of aforesaid comment it is interesting
to see what $F$ may be
chosen given $T=1$, or $T=H$ (these are corresponding to ordinary
$t$-time and $\alpha $-time ($\alpha =\log a $)
respectively)\@. Other functions $T(\varphi )$
seems not to be so important.

If we initially have chaotic scenario, i.\ e.\  $V(\varphi ) >> V_{0}
$,
then the proper $F(\varphi )$
are (see (\ref{eq10}) )

\begin{equation}
  T=1 \; \longleftrightarrow \; \Fe{} \propto H^{-3/2}
\label{eq12a}
\end{equation}

\begin{equation}
  T=H \; \longleftrightarrow \; \Fe{} \propto H^{-1}
\label{eq12b}
\end{equation}

Some interesting examples of such equivalence between
various theories are presented in Table \ref{tb}
(note that we work in dimensionless units
keeping carefully only field dependence of listed functions)\@.

\begin{table}[t]
\phantom{.}
\vspace{1cm}
\begin{tabular}{||c|c|c|c|c|c|c|c||}
 \hline \hline
 {\cal N} & Model   & Old $V(\varphi)$ & $\Fe{}$   & $\Te{}$
  & ${\cal F}(\varphi)$ & New $\tV{}{} $             & Model  \\
\hline
  1       & Chaotic & \Xl{m}{2}        & \pr{-3/2} &     1
  & \pr{-1/2}           & $V_{0}-\XL{\gamma}{4}  $   & New    \\
\hline
  2       & Chaotic & \Xl{\lambda}{4}  & \pr{-3}   &     1
  & \pr{-2}             & $V_{0}-\XL{\mu}{2}     $   & New    \\
\hline
  3       & Chaotic & \Xl{\lambda}{4}  & \pr{-2}   & \mbox{$H$}
  & \pr{-1}             & $V_{0}-\XL{\gamma}{4}  $   & New    \\
\hline
  4       & Power Law & \kep{}      & \pro{\kep{-\ah}} & \mbox{$H$}
  & \pro{\kep{-\ah}}    & $V_{0}-\XL{\mu}{2}     $   & New    \\
\hline
 \end{tabular}

\caption[]{Some examples of equivalence. Each row represents
two equivalent models, the old model potentials listed in the
third column, the new ones in the column 7.\@
The field and time redefinitions are presented in the columnes 4-6
respectively. \label{tb} }

\end{table}

Equivalent theories must have equivalent predictions in their
equivalence range. Of course, we would not like to say that new and
chaotic inflationary models are entirely equivalent. They are
very different from the point of view of their global
characteristics including the structure of inflationary
Universe at extremely large scales (much larger then our
present horizon)\@. But this is in fact the viewpoint of God
who can see all the Universe simultaneously. To be modest,
let us accept a viewpoint of a human being who can see only
things inside his horizon. We call him a local observer.

The effect of inflationary era on our part of the Universe
is twofold. First, it has a number of destructive effects:
inflation almost entirely prevents appearance of primordial
monopoles, gravitinos, domain walls, deviations from isotropy,
homogeneity, flatness, {\em etc}\@. All of this is simply
due to the rapid growth of the scale factor of the Universe
and do not depend on details of inflationary era. On the
other hand, inflation has some constructive effects, too.
That is the generation of primordial energy density
fluctuations (PEDF) responsible for the subsequent
formation of the visible large scale structure of the
Universe. Almost all things observable at cosmological
scales are related to PEDF (some of them are related to PEDF
immediately, others are expressed in terms of their first
and second derivatives)\@. Namely

\begin{description}
\item[1)] the fluctuations of the microwave background radiation
are related to the PEDF,
\item[2)] the peculiar velocities of galaxies are related to
gradients
of the newtonian potential which is proportional to PEDF,
\item[3)] the large scale structure of the matter distribution in the
visible part of the Universe and galaxy-galaxy and cluster-claster
correlation functions are proportional to the Laplasian of PEDF.
\end{description}
Therefore, theories which are
equivalent from the viewpoint of local observer must
produce the same spectrum of PEDF\@. Let us examine
our equivalence relations using this criterion. As it was
obtained in \cite{r5} (see also \cite{r2} for more
details and references) PEDF spectrum can be written
in the case of adiabatic fluctuations as

\begin{equation}
\frac{\delta \rho(k)}{\rho } = c \, \left.
\frac{\Vp{3/2}{}}{\Vps{}{\varphi }} \, \right|_{k \sim H(\varphi )}
\label{14.1}
\end{equation}
with constant $c$ depending on the details of Hot Big Bang. We
assume, that postinflationary eras in equivalent theories are
similar, so

\begin{equation}
\left. \frac{\Vp{3/2}{}}{\Vps{}{\varphi }} \, \right|_{k \sim
H(\varphi )} =
\left. \frac{\tV{3/2}{}}{\tVs{}{{\cal F}}} \, \right|_{k \sim
\tilde{H}({\cal
F})}
\label{eq14.2}
\end{equation}

Since we want to have the same value of PEDF spectrum at the same
wavelengths, the following constraint should be imposed (note, that
in (\ref{eq14.2}) and further $\varphi \neq \varphi({\cal F})$,
it is some value of the inflaton field in an old model, whereas
${\cal F}$ is some value of the inflaton field in a new one)

\begin{equation}
H(\varphi ) = H({\cal F})  \label{eq14.3}
\end{equation}
and, therefore,

\begin{equation}
\Vp{}{} = \tV{}{}  \label{eq14.4}
\end{equation}
for values of $\varphi $ and ${\cal F}$ of the old and new models,
for which the Eq.\  (\ref{eq14.2}) is valid. Thus we obtain
(the sign is unimportant):

\begin{equation}
\Vps{}{\varphi } = \left. \tVs{}{{\cal F}} \,
\right|_{\Vp{}{} = \tV{}{} }  \label{eq14.5}
\end{equation}

An easy way to see that this is indeed the case when relating
the chaotic and the new scenarios is to propose that at the very end
of inflation (when the PEDF with wavelengths that are inside
of our present horizon were produced) the value of inflaton field
potential energy $\Vp{}{} = \tV{}{} $ is somewhat smaller, then
$V_{0}$
(remember, that $V_{0}$ is the symmetry restoration energy
in the new inflationary model, as it can be seen from (\ref{eq11})
)\@.
Then at that range of inflaton dynamics one has (assuming the
sufficiently smooth time reparametrization)

\begin{equation}
F \sim 1 \; \Rightarrow \;
\mbox{$\tilde{V}$}({\cal F}_{e}) \sim V(\varphi_{e})  \label{eq14.6}
\end{equation}
and (\ref{eq14.5}) follows immediately. Here the subscripts ``e''
denote the values of inflaton fields near the ``end of inflation''
boundary. It is worth noting that the above proposition is valid
in almost all inflationary models \cite{r2} \@.

We have more to say about the notion of local observer and its
consequences for cosmology and, in particular, for the interpretation
of the presented equivalence relations. Let us postpone this
discussion to the end of this paper and proceed now to another
issue, intimately related with the relations (\ref{eq9}),
(\ref{eq10})\@.

Recently it was observed \cite{r7} ,
that the $\lambda \varphi^{4} $
theory is solvable in the slow rolling approximation, i.e.\  the
probability density distribution $P(\varphi ,t)$
can be found in the closed form. One can see from the second row in
the Table \ref{tb} that this theory is equivalent to the simplest
solvable new inflationary model
$\tV{}{} = V_{0} - \frac{1}{2} \mu^{2} {\cal F}^{2} $ \@.
Thus, the solvability of this theory is evident in our approach.
Moreover, we shall show later that all theories, listed in
Table \ref{tb}, are solvable in that sense.

Solvability in the slow rolling approximation means that Eq.\
(\ref{eq3})
can be reduced to the linear one by some field and time
redefinitions.
If so, the corresponding Fokker-Planck equation can be reduced to the
time-dependent Schr\"odinger equation for oscillator or for Coulomb
potential. Let us present the criterion of reducibility and give
the most interesting potentials satisfying this criterion. Suppose
for
a moment that (\ref{eq3}) is reduced, i.e.\  we have
\begin{equation}
   \frac{FV'_{\varphi }}{V^{1/2} T}=\kappa_{1}{\cal F}+\kappa_{2}
\label{eq13a}
\end{equation}
\begin{equation}
   \frac{FV^{3/4}}{T^{1/2}}=\theta_{1} {\cal F} + \theta_{2}
\label{eq13b}
\end{equation}
Note that in the case $\theta_{1}=0$,
the corresponding stochastic process ${\cal F}(\tau )$
remains gaussian, while in the case $\theta_{1} \neq 0$
it becomes nongaussian.

After some easy manipulations one can derive from (\ref{eq13a})
,(\ref{eq13b})
\begin{equation}
 \frac{\Vp{3/4}{}}{\Te{1/2}} \left(\frac{\Vps{}{\varphi }}{\Vp{5/4}{}
\,
 \Te{1/2}}\right)'_{\varphi } = \theta_{1}
 \frac{\Vps{}{\varphi }}{\Vp{5/4}{} \,
 \Te{1/2}} + \kappa_{1}
\label{eq14}
\end{equation}
and the complete (Gikhman-Skorokhod) criterion is
\begin{equation}
 \left(\frac{1}{\left(\frac{\Vps{}{\varphi }}{\Vp{5/4}{} \,
 \Te{1/2}}\right)'_{\varphi }}
 \left(\frac{\Vp{3/4}{}}{\Te{1/2}}
 \left(\frac{\Vps{}{\varphi }}{\Vp{5/4}{} \,
  \Te{1/2}}\right)'_{\varphi }\right)'_{\varphi }\right)'_{\varphi }
  = 0          \label{eq15}
\end{equation}
The criterion in the case $\theta_{1}=0$ is more simple
\begin{equation}
\left(\frac{\Vp{3/4}{}}{\Te{1/2}}\left(
\frac{\Vps{}{\varphi }}{\Vp{5/4}{} \,
\Te{1/2}}\right)'_{\varphi }\right)'_{\varphi } = 0
\label{eq16}
\end{equation}
The interpretation of the equations (\ref{eq15}), (\ref{eq16})
is twofold. One can say that
\begin{description}
\item[a)]   the potential $V(\varphi)$
is given and the Eqs.\ (\ref{eq15}) ,(\ref{eq16})
determine the corresponding time parameterization $T(\varphi)$,
where this model is solvable;
\item[b)] the time parameterization is given and the Eqs.\
(\ref{eq15}) ,(\ref{eq16})
determine the class of solvable potentials $V(\varphi)$ \@.
\end{description}

{}From the viewpoint {\bf a)} in the case $\theta_{1}=0$
we derive from (\ref{eq16})
\begin{equation}
  \Te{} = \frac{(\Vps{}{\varphi })^{2}}{C_{1}
  \Vp{3/2}{} + C_{2} \Vp{5/2}{}}
                                  \label{eq17}
\end{equation}

{}From the viewpoint {\bf b)} we should first set $T(\varphi)$\@.
As it was already noticed, the important cases are $T=1$, $T=H$\@. If
$T=1$, the only interesting solution of (\ref{eq17})
is ($C_{2}=0$)
 \[ \Vp{}{}=\frac{C_{1}^{2}}{2^{8}} (\varphi -\varphi_{0})^{4}  \]
which is of chaotic type. If $T=H$, there are the following
interesting
solutions (in the first example $C_{2}=0$\,)
 \[ \Vp{}{}=V_{0} e^{\sqrt{C_{1}} \varphi}  \]
 \[ \Vp{}{}=\frac{C_{1}}{C_{2}}
\frac{1}{\cosh^{2}(\frac{\sqrt{C_{1}}}{2}
   (\varphi -\varphi_{0}))}  \]
The former is of chaotic type, while the latter is of new
inflationary
type (note, however, that the second potential  is  in
fact indistinguishable
from $\tV{}{} =V_{0}- \frac{1}{2} \mu^{2} {\cal F}^{2}$
in the framework of the new inflationary scenario)\@.

Now let us pass to the $\theta_{1} \neq 0$
case. Within the {\bf b)} interpretation one can derive
from (\ref{eq14}) in the case $T=H$
\begin{equation}
  u''_{\varphi \varphi}=\theta_{1} u u'_{\varphi } - \kappa_{1} u
\label{eq18}
\end{equation}
where $u(\varphi) = \Vp{-1/2}{} $\@.
The parametric form of the solution of Eq.\ (\ref{eq18})
is available, but we present here only one interesting example
 \[ \Vp{}{}= \frac{\theta^{2}_{1}}{4} (\varphi -\varphi_{0})^{2} \]
(but \(\mbox{\Vp{1/2}{}}=-\frac{\theta_{1}}{2} (\varphi
-\varphi_{0})\),
because $\theta_{1}<0$, $\kappa_{1}=0$ )\@.

In the case $T=1$, $\theta_{1} \neq 0$
we have not found any interesting solution of the corresponding
equation
\begin{equation}
  w''_{\varphi \varphi}=\theta_{1} w^{3} w' -\kappa_{1} w^{3}
\label{eq19}
\end{equation}
where $w=\mbox{\Vp{1/4}{}}$ \@.

Let us now return to the equivalence relations. Should we say
hereafter,
that one must treat both the new and the chaotic inflationary
scenarios
on equal footing? Is there the logical
basis to identify them, although their global (invisible!\ )
properties
are so different. To answer this questions, one has to elaborate some
sort of strategy (or better to say methodology) for cosmology, since
this
branch of knowledge cannot be called now the natural science in the
sense usually applied to terrestrial physics. We hope that Sakharov
Conference is just right place to present our approach to such an
important subject.

The heart of our approach is the notion of ``correlational
predictions''. In the course of the previous
development of physics always there was an easy framework to
ensure both the predictive and the objective character of
natural science. That is, before every physicist had clear
understanding that things, which he deals with, are available
at least after some known indirect measurements, if not immediately.
Another point was that he always was sure that he can examine the
constructed theory by a new experiment and accept or reject it
after this test. Now the situation is rather different.
We don't know what the whole Universe is and we have no chance
to see its global view, because we are mortal. At the same time
we have no chance to see another ``experiment'', because we live
inside of this particular ``experiment''. Nevertheless, we believe
that The Nature is knowable to much more extent, than the limits
of our visible part of the Universe.

Indeed, what does the ``correlational prediction'' mean?
One should make the set of the cosmological (astrophysical)
observations, then go to the theoretical explanation of the observed
phenomena, then make new predictions on the basis of constructed
theory and finally look for existence of newly predicted
phenomena. This is principally different from the
usual terrestrial framework, because the predictions are
highly correlated with what was observed previously.
They are in fact the ``functions'' of those observations
only. In usual situation
one always has some additional information about
the system under investigation like boundary conditions,
statistics of experiments in various environments, etc.
But instead of this lack of knowledge (and due to it)
one has in cosmology a good chance to probe the
principally unobservable things, including them
in the theoretical formalism, used to make the correlational
predictions. In such a case the better fit of correlations
between observed quantities means the better choice of
description of (otherwise unobservable) things.
One should note the similarity between this prescription
and the familiar weak anthropic principle. But we think that
the existence of the mankind is no more important than the
existence of all other observable things in the visible part of the
Universe. So, our prescription simply states some very natural
generalization of the weak anthropic principle in order
to include all observables in the set of quantities,
upon which the joint probability
distribution of new and old observations depends.
The answers in cosmology should look like the following:

\begin{quote}
{\em Provided that we have observed the (list of observations),
we should think,  that  (new  prediction)   also   take   place.
Otherwise our theoretical paradigm, predicting the high
correlation between the new prediction and old observations is
false.}
\end{quote}

It is worth noting that the number of observable things
inside of our horizon is so huge, that one can hope to have
a job even in centuries. The other important point is that
our prescription makes ``paradigms'', or ``scenarios'' the
only reasonable framework to investigate the Universe. This is
the fundamental limitation, rather than technical.

Accepting this approach, we can obtain
some consequences from the equivalence of different
inflationary scenarios. Indeed, the new (or variant thereof) scenario
could be favored against the chaotic one (and its relatives)
due to more natural emergence of the inflaton potential
of the former type in the low energy effective field
theory like the standard model. On the other hand,
chaotic scenarios have many properties that make them
more natural if we see the global (unobservable)
structure of inflationary Universe. Since we can say
now that both types of scenarios are equivalent from
the viewpoint of local observer, the
advantages of the chaotic scenario becomes crucial.
That's why we think that it is favored by the notion of
equivalence.

\end{document}